\documentclass[a4paper,11pt]{article}
\pdfoutput=1 
\usepackage{jcappub} 
\usepackage[T1]{fontenc} 
\usepackage[utf8]{inputenc}
\usepackage{nicefrac}
\usepackage{amsfonts}
\usepackage{lmodern}
\usepackage{amsthm,graphicx}
\usepackage{epsfig}
\usepackage[normalem]{ulem}
\usepackage{latexsym,amssymb} 
\usepackage{amsmath,mathtools}
\usepackage{xcolor,xspace}
\usepackage{siunitx}
\usepackage{multirow}
\usepackage[font=footnotesize, labelfont=bf]{subcaption}
\usepackage{booktabs} 
\usepackage{empheq}
\usepackage{subcaption}
\usepackage{aas_macros}
\usepackage{longtable}
\usepackage{titlesec}

\arxivnumber{2411.02774}

\title{\boldmath Non-Gaussianity in CMB lensing from full-sky simulations}

\author{Jan Hamann}
\author[1]{and Yuqi Kang\note{Corresponding author.}}

\affiliation{Sydney Consortium for Particle Physics and Cosmology, \\School of Physics, The University
of New South Wales, \\Sydney NSW 2052, Australia}

\emailAdd{jan.hamann@unsw.edu.au}
\emailAdd{yuqi.kang@unsw.edu.au}

\abstract{The lensing convergence field describing the weak lensing effect of the Cosmic Microwave Background (CMB) radiation is expected to be subject to mild deviations from Gaussianity.  We perform a suite of full-sky lensing simulations using ray tracing through multiple lens planes – generated by combining $N$-body simulations on smaller scales and low-to-intermediate redshifts with realisations of Gaussian random fields on large scales and at high redshifts.  We quantify the non-Gaussianity of the resulting convergence fields in terms of a set of skewness and kurtosis parameters and show that the non-Gaussian information in these maps can be used to constrain cosmological parameters such as the cold dark matter density $\Omega_\mathrm{c} h^2$ or the amplitude of primordial curvature perturbations $A_\mathrm{s}$. We forecast that for future CMB lensing observations, combining the non-Gaussian parameters with the Gaussian information can increase constraining power on $(\Omega_\mathrm{c} h^2, A_\mathrm{s})$ by a few per cent compared to constraints from Gaussian observables alone. We make the simulation code for the full-sky lensing simulation available for download from \href{https://github.com/Kang-Yuqi/FLAReS}{GitHub}.\footnote{\href{https://github.com/Kang-Yuqi/FLAReS}{\texttt{https://github.com/Kang-Yuqi/FLAReS}}}}

\begin{document}
\maketitle
\flushbottom

\section{Introduction}
    On their way from the surface of last scattering to us, the photons of the Cosmic Microwave Background (CMB) are subject to small deflections due to gravitational lensing caused by the large scale structure of the Universe.  This leads to slight distortions in observed CMB temperature and polarisation maps, which carry information about the matter distribution at lower-redshift that can be accessed via various reconstruction methods (e.g.,~\cite{Okamoto:2003zw, Kesden:2003cc, Carron:2017mqf, Millea:2020cpw, Floss:2024gqk}).  Weak gravitational lensing of the CMB was detected at high statistical significance in \textit{Planck} data~\cite{Planck:2018lbu} and high-resolution ground-based CMB polarisation experiments like AdvACT~\cite{ACT:2023kun}, SPT-3G~\cite{SPT:2023jql}, and the upcoming Simons Observatory~\cite{SimonsObservatory:2018koc} and CMB-S4~\cite{CMB-S4:2016ple} experiments can provide even more detailed CMB lensing measurements.

    Due to its ability to probe the total matter distribution over a broad range of redshifts, CMB weak lensing  has become a staple tool for the inference of cosmological parameters.  At the technical level, this usually involves a likelihood analysis based on the power spectrum of maps of the lensing potential – an approach that is limited to extracting the map's Gaussian information.  On the other hand, CMB lensing maps are expected to contain a non-trivial non-Gaussian contribution due to the inherent non-linearity of gravitational lensing itself~\cite{Lewis:2011fk}, and the fact that the lens (i.e., matter perturbations at lower redshift) is non-Gaussian.  The resulting non-Gaussianity of the lensing map is dependent on cosmological parameters, and therefore becomes a useful probe of cosmology in itself.
    While for current data a focus on the power spectrum is indeed perfectly appropriate given that the non-Gaussianity of the \textit{Planck} lensing map is dominated by reconstruction noise~\cite{Hamann:2023tdu}, in the future, the non-Gaussian information will become observable~\cite{Liu:2016nfs,Namikawa:2016jff,Zhang:2022wfi,Kalaja:2022xhi} and can be used to complement the power spectrum in parameter inference.

    In order to actually access this information, one also requires accurate theoretical predictions of non-Gaussian lensing observables though.  The problem's non-linearity necessitates a combination of $N$-body simulations to model the matter perturbations with ray-tracing techniques.  This approach has been applied to the weak gravitational lensing of galaxies~\cite{Fosalba:2007mf, Fosalba:2013mra, Wei:2018uwb, Ferlito:2024gmi} as well as the CMB~\cite{Carbone:2007yy, Das:2007eu, Calabrese:2014gla, Takahashi:2017hjr, Liu:2017now, Stein:2020its}.

    In this work, we present a simulation pipeline for the full-sky CMB lensing convergence maps, with an emphasis on their non-Gaussian properties.  We keep the computational requirements manageable by employing multi-scale $N$-body simulations on small angular scales and low-to-intermediate redshifts, and resorting to realisations of Gaussian random fields on large scales and at high enough redshifts.  After validating our method and tuning some of its input parameters, we apply the results of our simulations to estimate to what extent the non-Gaussian information can improve cosmological parameter constraints compared to the Gaussian case.
    
    The paper is structured as follows. In Section~\ref{sec:CMB lensing}, we describe our lensing simulation algorithm. In Section~\ref{sec:validation} we validate the simulation results.  We forecast the potential of non-Gaussian information to improve constraints on cosmological parameters in Section~\ref{sec:constrain} and present our conclusions in Section~\ref{sec:conclusion}.

\section{CMB lensing simulations}\label{sec:CMB lensing}
    Let us start this Section with a brief summary of the main results related to weak lensing relevant for the discussion of this paper.  For more detailed reviews of the weak lensing formalism, we refer the reader to, e.g., Refs.~\cite{Hu:2000ee, Bartelmann:1999yn,Refregier:2003ct,Lewis:2006fu}.
    
    The path of a photon traveling a short distance $\delta r$ through a gravitational potential $\Psi$ is subject to a local deflection by an angle $\delta \beta = -\dfrac{2}{c^2}\delta r \, \nabla_{\perp}\Psi$, where $\nabla_{\perp}$ denotes the gradient in the transverse direction.  To an observer located at $r=0$, a source at comoving distance $r_s$ subjected to a local deflection $\delta \beta \ll 1$ at distance $r$ will appear to be shifted with respect to its true position by an observed local deflection angle $\delta \theta = \dfrac{r_s-r}{r_s}\delta \beta$.  
    The total observed deflection angle $\alpha$ can then be obtained by integrating over $\delta \theta$ along the photon path.  In the Born approximation, one instead integrates along the line of sight:
    \begin{equation}\label{eq:deflection angle}
    \alpha^\mathrm{B} = -2 \int_0^{r_s} \mathrm{d}r ~\frac{r_s-r}{r_s c^2} \nabla_{\perp}\Psi(r \hat{\mathbf{n}}, r).
    \end{equation}
    
    The observable we consider here is the lensing convergence $\kappa$ which is related to the deflection angle by $\kappa = -\frac{1}{2}\nabla_{\hat{\mathbf{n}}}\alpha$, where $\nabla_{\hat{\mathbf{n}}}$ is the covariant derivative on the sphere defined by $\hat{\mathbf{n}}$. Equation~\ref{eq:deflection angle}, combined with the relation $\nabla_{\perp}\Psi = \frac{1}{r}\nabla_{\hat{\mathbf{n}}}\Psi$ then leads to the following expression for $\kappa$ in the Born approximation:
    \begin{equation}\label{eq:kappa_from_potential}
    \kappa^\mathrm{B} = \int_0^{r_s} \mathrm{d}r ~\frac{(r_s-r)r}{r_s c^2} \nabla^2_{\hat{\mathbf{n}}}\Psi(r \hat{\mathbf{n}}, r).
    \end{equation}
    
    The gravitational potential $\Psi$ is related to the density contrast $\delta=\frac{\rho-\bar{\rho}}{\bar{\rho}}$ via Poisson's equation,
    \begin{equation}\label{eq:lensing_potential_density}
    \nabla^2 \Psi = 4\pi G\rho a^2 =  \frac{3H_0^2 \Omega_\mathrm{m} }{2 a c^2} \delta,
    \end{equation}
    where $\Omega_\mathrm{m}$ is the total matter density, $H_0$ is the Hubble constant and $a$ is the scale factor.  Augmenting the spherical Laplacian $\nabla_{\hat{\mathbf{n}}}^2$ to the 3-dimensional Laplacian $\nabla^2$, and noting that the derivatives of $\Psi$ along the light path average to zero~\cite{Bartelmann:1999yn}, the lensing convergence can be expressed as

    \begin{equation}\label{eq:kappa_born}
    \kappa^\mathrm{B}(\hat{\mathrm{n}})=\frac{3 H_0^2 \Omega_\mathrm{m}}{2 c^2} \int_0^{r_s} \mathrm{d}r ~\delta(r, \hat{\mathrm{n}}) \frac{\left(r_s-r\right) r}{r_s a(r)}.
    \end{equation}

    While the Born approximation may in practice be sufficiently accurate for simulations of the power spectrum at low source redshifts, even for upcoming large lensing surveys, it has been shown that non-Gaussian observables will require post-Born corrections~\cite{Dodelson:2005rf,Hilbert:2008kb,Hagstotz:2014qea,Pratten:2016dsm,Petri:2016qya} to avoid biases. Since the accuracy of the Born approximation decreases with increasing source redshift and we are interested in weak lensing of the CMB, post-Born corrections of higher order in $\Psi$ will have to be taken into account.  However, merely going to second order is not sufficient due to the presence of significant partial cancellations between second- and third-order terms~\cite{Pratten:2016dsm}, so we consider post-Born corrections to the convergence up to third order in $\Psi$:    
    \begin{equation}\label{eq:kappa_3rdorder}
    \kappa = \kappa^\mathrm{B} + \kappa^{(2)} + \kappa^{(3)} + \mathcal{O}(\Psi^4). 
    \end{equation}
    The corresponding correction terms were derived in Ref.~\cite{Pratten:2016dsm} and, with the short-hand notation $\Psi_{,a} \equiv \nabla_{\perp}\Psi$ for transverse derivatives, read, in our notation:

    \begin{enumerate}
    \item{Second Order}
    \begin{equation}
    \kappa^{(2)}=\kappa^{2L}+\kappa^{2D}
    \end{equation}
    \begin{equation}
    \kappa^{2L}=-2 \int_0^{r_s} \mathrm{d} r \int_0^r \mathrm{d} r^{\prime}~r r^{\prime}  ~\frac{r_s-r}{r_s}  \frac{r-r^{\prime}}{r} \left[\Psi_{, a c}(r) \Psi_{, c a}\left(r^{\prime}\right)\right]
    \end{equation}
    \begin{equation}
    \kappa^{2D}=-\frac{3H_0^2 \Omega_\mathrm{m} }{c^2} \int_0^{r_s} \mathrm{d} r\int_0^r \mathrm{d} r^{\prime}~ \frac{r^2}{a(r)} ~\frac{r_s-r}{r_s}  \frac{r-r^{\prime}}{r} \left[\delta_{,c}(r) ~\Psi_{, c}\left(r^{\prime}\right)\right]
    \end{equation}
    \item{Third Order}
    \begin{equation}
    \kappa^{(3)}=\kappa^{3D}+\kappa^{3X}+\kappa^{3L}+\kappa^{3X1}+\kappa^{3X2}
    \end{equation}
    \begin{equation}
    \kappa^{3D}=\frac{3H_0^2 \Omega_\mathrm{m} }{c^2}\int_0^{r_s} \mathrm{d} r \int_0^r \mathrm{d} r^{\prime} \int_0^r \mathrm{d} r^{\prime \prime}  \frac{r^3}{a(r)}~\frac{r_s-r}{r_s}  \frac{r-r^{\prime}}{r} \frac{r-r^{\prime\prime}}{r} \left[\delta_{,cd}(r) \Psi_{, c}\left(r^{\prime}\right) \Psi_{, d}\left(r^{\prime \prime}\right)\right]
    \end{equation}
    \begin{equation}
    \kappa^{3X}=4 \int_0^{r_s} \mathrm{d} r \int_0^r \mathrm{d} r^{\prime} \int_0^r \mathrm{d} r^{\prime \prime}~r^2 r^{\prime}\frac{r_s-r}{r_s}  \frac{r-r^{\prime}}{r} \frac{r-r^{\prime\prime}}{r} \left[\Psi_{, a c d}(r) \Psi_{, a c}\left(r^{\prime}\right) \Psi_{, d}\left(r^{\prime \prime}\right)\right]
    \end{equation}
    \begin{equation}
    \kappa^{3L}=4 \int_0^{r_s} \mathrm{d} r \int_0^r \mathrm{d} r^{\prime} \int_0^{r^{\prime}} \mathrm{d} r^{\prime \prime} ~r r^{\prime} r^{\prime \prime}~\frac{r_s-r}{r_s}  \frac{r-r^{\prime}}{r} \frac{r^{\prime}-r^{\prime\prime}}{r^{\prime}} \left[\Psi_{, c d}(r) \Psi_{, ac}\left(r^{\prime}\right) \Psi_{, a d}\left(r^{\prime \prime}\right)\right]
    \end{equation}
    \begin{equation}
    \kappa^{3X1}=\frac{6H_0^2 \Omega_\mathrm{m} }{c^2}\int_0^{r_s} \mathrm{d} r \int_0^r \mathrm{d} r^{\prime} \int_0^{r^{\prime}} \mathrm{d} r^{\prime \prime}~\frac{r^2 r^{\prime}}{a(r)} ~\frac{r_s-r}{r_s}  \frac{r-r^{\prime}}{r} \frac{r^{\prime}-r^{\prime\prime}}{r^{\prime}}~\left[\Psi_{, d}\left(r^{\prime \prime}\right) \Psi_{, c d}\left(r^{\prime}\right) \delta_{, c}(r)\right]
    \end{equation}
    \begin{equation}\label{eq:kappa_3X2}
    \kappa^{3X2}=4 \int_0^{r_s} \mathrm{d} r \int_0^r \mathrm{d} r^{\prime} \int_0^{r^{\prime}} \mathrm{d} r^{\prime \prime}~r r^{\prime 2}~\frac{r_s-r}{r_s}  \frac{r-r^{\prime}}{r} \frac{r^{\prime}-r^{\prime\prime}}{r^{\prime}}~\left[\Psi_{, d}\left(r^{\prime \prime}\right) \Psi_{, a c}(r) \Psi_{, a c d}\left(r^{\prime}\right)\right]
    \end{equation}
    \end{enumerate}

    For the purpose of numerically simulating convergence maps, we approximate the lensing process by a series of lensing events on thin lenses.  This approach is known as multiple-plane weak lensing simulation~\cite{Jain:1999ir} and has been adopted in several previous studies~\cite{Carbone:2007yy,Fosalba:2007mf,Das:2007eu,Hilbert:2008kb,Teyssier:2008zd,Takahashi:2017hjr,Wei:2018uwb,Ferlito:2023gum}.
    
    In this case, the integrals in the above expressions are replaced by discrete sums over $N$ spherical shells of thickness $\Delta r_j$, centered on the observer. For the Born-approximated convergence (Equation~\eqref{eq:kappa_born}), one obtains
    \begin{equation}\label{eq:kappa_numerical}
    \kappa^\mathrm{B}(\hat{\mathrm{n}}) \simeq \frac{3 H_0^2 \Omega_\mathrm{m}}{2 c^2} \sum_{j=1}^N \hat{\delta}(r_j,\hat{\mathrm{n}}) \frac{\left(r_s-r_j\right) r_j}{r_s a_j} \Delta r_j,
    \end{equation}
    where $j$ is the index of the shell, $r_j$ denotes its effective radius (see Section~\ref{sec:shell_construction}) and $\hat{\delta}$ is the radial projection of the density contrast onto a sphere of radius $r_j$.  The corresponding expressions for the post-Born corrections are constructed in the same way.

    In the limits of high enough distance/redshift or on large enough scales, the underlying matter field is still well within the regime of linear evolution, and in the absence of a substantial primordial non-Gaussianity, $\delta$ and $\Psi$ will be well-approximated by Gaussian random fields whose power spectra can be calculated straightforwardly within linear perturbation theory.  
    
    On smaller scales and at lower redshifts however, the matter field will have started undergoing non-linear evolution and one will have to resort to $N$-body simulations to predict $\delta$ and $\Psi$.  Given that our goal is to simulate the full sky and since the CMB lensing kernel covers a large range of redshifts, requiring simulations from $0 < z < \mathcal{O}(10)$, maintaining a reasonable angular resolution would necessitate a huge simulation if one were to stick to a single resolution over the entire volume~(e.g., \cite{Carbone:2007yy}).  
    In the interest of keeping the computational load manageable, we therefore adopt a multi-resolution approach as in Ref.~\cite{Takahashi:2017hjr}, and construct maps of the matter field at different distances from $N$-body simulations run at different physical resolutions.  In practice, we keep the particle numbers of our simulations fixed, but use increasingly larger simulation box sizes as the distance from the observer increases.

\subsection{Reference cosmology and geometrical arrangement of lensing shells \label{sec:shell_arrangement}}

    Unless explicitly mentioned otherwise, we will adopt as our reference cosmology the base $\Lambda$CDM model with parameter values $\Omega_\mathrm{b}h^{2} = 0.02216$, $\Omega_\mathrm{cdm} h^{2} = 0.1203$, $A_\mathrm{s} = 2.119\times10^{9}$, $n_\mathrm{s} = 0.96$ and $h = 0.67$ – identical to the settings used for generating the \textit{Planck} FFP10 simulations~\cite{Planck:2018lkk}.

    \begin{figure}[htbp]
        \centering
    	\includegraphics[width=0.75\textwidth]{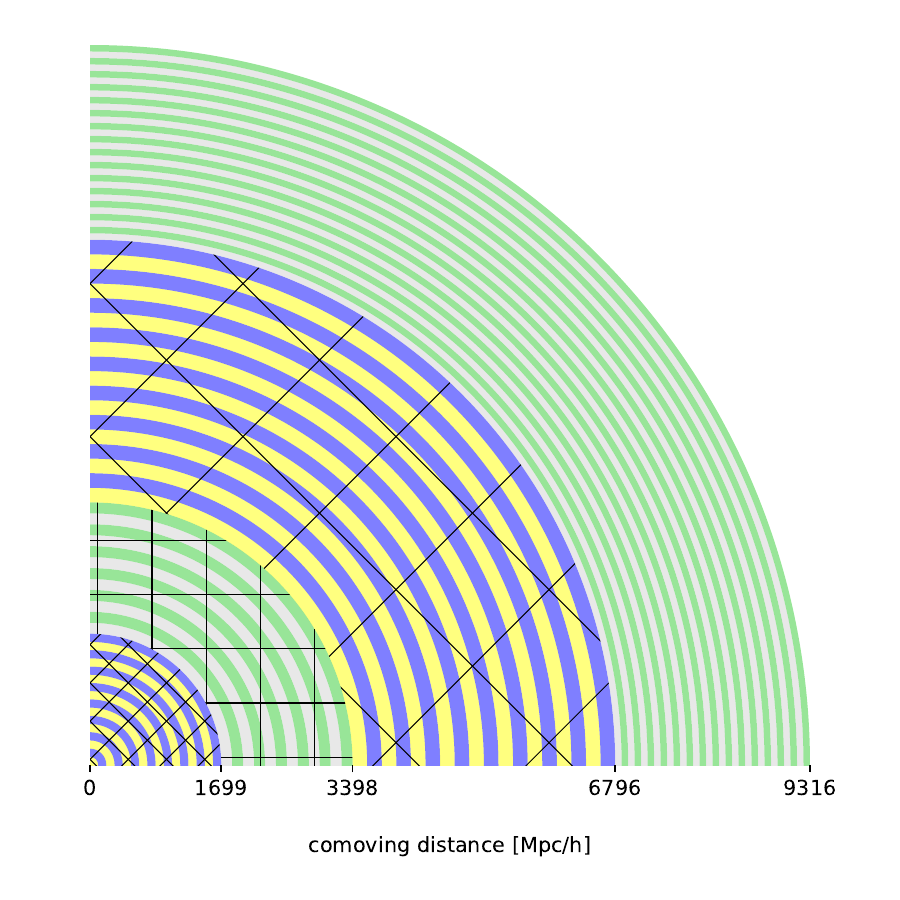}
    	\caption{Schematic of the geometric arrangement of lensing shells, presented as a quarter cross-sectional view. The grid represents the stacking of N-body simulation boxes, while the colored rings indicate the locations of the lensing shells. Different color combinations are used to represent variations in shell thickness. Both the box size and shell thickness are shown to scale.\label{fig:nbody_arrange}}
    \end{figure}

    We sketch our default geometrical arrangement of lensing shells in Figure~\ref{fig:nbody_arrange}.  Taking the observer to reside at comoving distance $r_0 = 0$, we fill the volume between $r_{i-1}$ and $r_i$ with $N_i$ spherical lensing shells of equal thickness $\Delta r_i = (r_{i} - r_{i-1})/N_i$ whose density fields are constructed from $N$-body simulations of box size $L_i$, with $i \in \left\{1, ..., n_\mathrm{box}\right\}$.  The distances $r_i$ are chosen such that at $r_i$, a box of size $L_i$ subtends an angle of approximately 11.6~degrees on the sky – roughly the angular scale associated to a multipole of $\ell \sim 15$.
    
    The remaining volume between $r_{n_\mathrm{box}}$ and the surface of last scattering at $r_s$ is sliced into $N_\mathrm{G}$ lensing shells in which the density fields are realisations of Gaussian random fields, generated from the angular power spectrum of matter perturbations at the corresponding distance.
    A summary of the settings used for construction of the lensing shells can be found in Table~\ref{tab:shell_settings}.  For these settings, the shell thicknesses are $\Delta r_i \in \{106.2, 141.6,188.8\} \, \rm{Mpc}/h$ for the $N$-body shells and $\Delta r = 84 \,\rm{Mpc}/h$ for the Gaussian shells.   In our reference cosmology, the redshifts associated with the $r_{i}$ are $z_i \in \{0.68, 1.83, 12.61\}$.

    \begin{table}[t]
    \centering
    \begin{tabular}{c|c|c}
        \hline \hline  Symbol & Parameter & Fiducial value \\ \hline
        $n_\mathrm{box}$ & number of different $N$-body box sizes & 3 \\
        $L_i$ & side length of $i$th $N$-body box & $\{350, 700, 1400\}$~Mpc/$h$ \\     
        $r_{i}$ & outer radius of largest lensing shell using box of size $L_i$  & $\{1699, 3398, 6796\}$~Mpc/$h$ \\
        $N_i$ & number of lensing shells using box of size $L_i$ & \{16, 12, 18\}\\
        $N_\mathrm{G}$ & number of Gaussian lensing shells between $r_{n_\mathrm{box}}$ and $r_s$ & 30 \\
        \hline
        $N_\mathrm{side}^\mathrm{lens}$ & \texttt{HEALPix} resolution of lens maps & 2048 \\
        \hline \hline
    \end{tabular}
    \caption{The parameters that define the geometry of lensing shells.  We refer the reader to Section~\ref{sec:validation} for more details about the choice of these parameters. \label{tab:shell_settings}}
    \end{table}

    \subsection{$N$-body Simulations}
        We use the $N$-body code \texttt{Gadget4}~\cite{Springel:2020plp} to perform dark matter only $N$-body simulations with $N = 512^3$~particles, imposing periodic boundary conditions on the cubical simulation volumes and setting the softening length to $4\%$ of the mean particle separation.  Initial conditions are set at redshift $z=120$ and generated based on second order Lagrangian perturbation theory (2LPT), with the initial linear power spectrum calculated using the Boltzmann solver \texttt{CAMB}~\cite{Lewis:1999bs,Howlett:2012mh}.  At each box size, we run \texttt{Gadget4} three times with different realisations of the initial conditions and output snapshots of the simulations at all redshifts corresponding to the effective radii (see below) of the respective lensing shells.

    \subsection{Construction of Density Contrast Maps \label{sec:shell_construction}}
        For a lensing shell with index $j$ whose inner and outer radii are $r_{\mathrm{in},j}$ and $r_{\mathrm{out},j}$, respectively, we place the lens plane at an effective radius given by the cone-volume-weighted mean distance~\cite{Shirasaki:2015dga,Takahashi:2017hjr},
        \begin{equation}\label{eq:shell_radius}
        \bar{r}_j = \frac{\int_{\Delta r}\mathrm{d}r~r r^2}{\int_{\Delta r}\mathrm{d}r~r^2} = \frac{3}{4} \frac{(r^4_{\text{out},j} - r^4_{\text{in},j})}{(r^3_{\text{out},j} - r^3_{\text{in},j})}.
        \end{equation}        
        Note that at large distances $r \gg \Delta r$, $\bar{r}_j$ converges to the arithmetic mean between $r_{\mathrm{in},j}$ and $r_{\mathrm{out},j}$.

        Lenses are represented by \texttt{HEALPix}~\cite{Gorski:2004by,Zonca2019} maps of the projected density contrast $\hat{\delta}$ at resolution $N_\mathrm{side}^\mathrm{lens} = 2048$.
    
        \subsubsection{Lensing shells at high redshifts}\label{sec:high z shells}
            
        For lensing shells with effective comoving radius $\bar{r}_j > r_{n_\mathrm{box}}$ (i.e., $z > 12.61$\footnote{ This turns out to be sufficiently large to capture most of the non-Gaussian information in CMB lensing, as we will show in Section~\ref{sec:validation}.}), we use the Boltzmann code \texttt{CAMB}~\cite{Lewis:1999bs,Howlett:2012mh} to compute the matter power spectrum $P_\delta(k, z(\bar{r}_j))$ including \texttt{Halofit}~\cite{Takahashi:2012em} non-linear corrections, and then calculate the angular power spectrum of the density contrast in the Limber approximation~\cite{1953ApJ...117..134L}\footnote{Note that at the largest scales this will be somewhat inaccurate and for an analysis of real data one should use the curved-sky expression instead.  However for $\ell > 10$, the Limber approximation is accurate at better than 1\%~\cite{Jeong:2009wi}.}
        \begin{equation}\label{eq:cl_delta_theory_limber}
            C_{\ell}^\delta (z(\bar{r}_j)) \simeq P_\delta\left(\frac{\ell+0.5}{\bar{r}_j}, z(\bar{r}_j)\right),
        \end{equation}
        which serves as input for the \texttt{HEALPix} subroutine \texttt{synfast} to generate a Gaussian realisation $\hat{\delta}^\mathrm{G}$ of the density contrast map.

        \subsubsection{Lensing shells at lower redshifts}
\begin{figure}[htbp]
            \centering
    	    \includegraphics[width=0.8\textwidth]{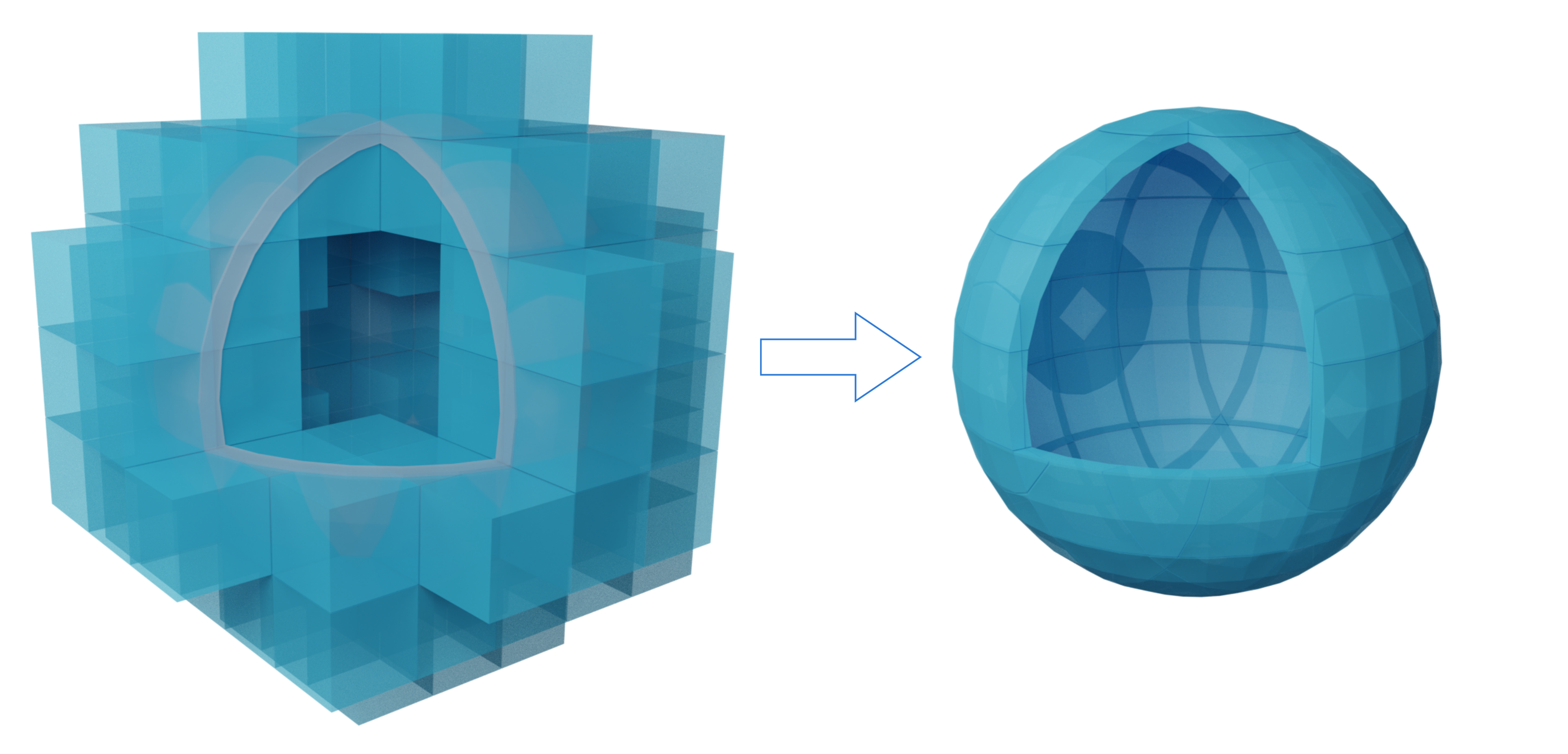}
    	    \caption{This illustration shows the stacking of $N$-body simulation boxes and the construction of a lensing shell.}
            \label{fig:lens_shell}
        \end{figure}

        Lensing shells with effective comoving radius $r_j < r_{n_\mathrm{box}}$ ($z < 12.61$) are constructed by
        stacking $N$-body boxes and slicing out the shell, as illustrated in Fig.~\ref{fig:lens_shell}. 
    
        We first project particles in each shell onto a \texttt{HEALPix} grid to obtain a map of the surface density $\sigma$ and then form a map of the projected density contrast $\hat{\delta}^\mathrm{sim} = \frac{\sigma-\bar{\sigma}}{\bar{\sigma}}$.  In our code, the processes of stacking boxes and projecting particles onto the pixelised sphere are performed on-the-fly and in parallel.

        As discussed in detail by Takahashi et al.~\cite{Takahashi:2017hjr}, one needs in addition to take into account the fact that the density contrast field is averaged over a shell of finite thickness which predominantly impacts fluctuations on scales larger than the shell thickness.  This effect can be corrected for by replacing the matter power spectrum by~\cite{Takahashi:2017hjr}
        \begin{equation}\label{eq:thickness_corr}
            P_\delta (k) \rightarrow \Delta r \, 
            \int \frac{\mathrm{d} k_{\|}}{2 \pi} P_\delta(k) \operatorname{sinc}^2\left(\frac{k_{\|} \Delta r}{2}\right),
        \end{equation}
        where $\operatorname{sinc}(x) \equiv \sin(x)/x$ and $k_{\|}$ is the projection of the wave vector $\mathbf{k}$ parallel to the line of sight.  Together with the perpendicular projection $\mathbf{k}_\perp$, we have $k=\left(k_{\|}^2+\left|\boldsymbol{k}_{\perp}\right|^2\right)^{1 / 2}$ with $\left|\boldsymbol{k}_{\perp}\right|=\ell/\bar{r}$.  Note that in the limit of an infinitely thick shell,
        \begin{equation}\label{eq:shell_power_spectrm}
        \lim_{k_{\|} \Delta r \rightarrow \infty} \Delta r \, 
            \int \frac{\mathrm{d} k_{\|}}{2 \pi} P_\delta(k) \operatorname{sinc}^2\left(\frac{k_{\|} \Delta r}{2}\right) = P_\delta (k).
        \end{equation}
        We compute the $\Delta r$- and $\ell$-dependent correction factors
        \begin{equation}
            \zeta(\Delta r, \ell = k \bar{r}) \equiv (P_\delta (k))^{-1} \, \int \frac{\mathrm{d} k_{\|}}{2 \pi} P_\delta(k) \operatorname{sinc}^2\left(\frac{k_{\|} \Delta r}{2}\right)
        \end{equation}
        and apply them at the level of the spherical harmonics coefficients $\delta_{\ell,m}^\mathrm{sim}$ of the density contrast map via the \texttt{HEALPix} filter routine \texttt{almfl}:
        \begin{equation}\label{eq:thickness_corr_delta}
        \hat{\delta}_{\ell,m}^\mathrm{sim, corr} = \sqrt{\zeta(\Delta r, \ell)} \, \delta_{\ell,m}^\mathrm{sim}.
        \end{equation}

        The density contrast fields constructed in this way obviously cannot model fluctuations on scales larger than the physical box size of the $N$-body simulations.  However, on sufficiently large scales, the fluctuations remain well-approximated by linear perturbation theory, and we can thus add fluctuations on larger angular scales in the form of a Gaussian realisation of the field, similar to the procedure at high redshifts.
        
        In practice, we construct the density contrast field by combining a high-pass filtered version of the density field constructed from the $N$-body simulations $\hat{\delta}^\mathrm{sim}$ with a low-pass filtered Gaussian realisation $\hat{\delta}^\mathrm{G}$:
        \begin{equation}
            \hat{\delta}_{\ell,m} = \Theta(\ell - \ell_{\mathrm{thr},j}) \hat{\delta}^\mathrm{sim, corr}_{\ell,m} + (1 - \Theta(\ell - \ell_{\mathrm{thr},j})) \hat{\delta}^\mathrm{G}_{\ell,m}.
        \end{equation}
        We determine the threshold multipole $\ell_{\mathrm{thr},j}$ for each shell by finding the $\ell$ corresponding to an angular separation given by the half the angle subtended on the sky by the respective $N$-body box.  For the second and third box size, we have $16 \leq \ell_\mathrm{thr} \leq 30$, whereas for the smallest box size $\ell_\mathrm{thr} \leq 30$ (the innermost shell fits completely into the smallest $N$-body box).  Thus the final convergence maps will have contributions from a mixture of Gaussian and $N$-body simulation shells at multipoles $\ell \leq 30$ and purely from $N$-body simulation shells at $\ell > 30$.

    \subsection{Construction of Convergence Maps}
        In order to avoid introducing spurious correlations between the structures in the shells populated from $N$-body simulations, we randomly select one of our three independent simulations for each shell.  In addition, we subject all of the lower redshift density contrast maps to random rotations.

        With the entire volume between the observer and the last scattering surface filled with lensing shells, and their respective maps of the projected density contrast constructed, we can now compute the corresponding maps of $\hat{\Psi}$ (via Poisson's equation) and evaluate their derivatives, which finally allows us to calculate the convergence map by computing the sums in Equations~\eqref{eq:kappa_born}-\eqref{eq:kappa_3X2}] for each pixel.  
        


\section{Tuning of simulation parameters and validation of simulations\label{sec:validation}}
    The simulation pipeline discussed in the previous Section requires a fair number of input parameters.  In this Section we will outline the considerations that went into the choice of simulation settings listed in Table~\ref{tab:shell_settings} – generally, these are guided by demanding consistency with analytical results where available and numerical convergence where not.  The former is relevant to the angular power spectrum of the convergence maps, and the latter is applied mostly to the analysis of our simulations' non-Gaussian properties.

\subsection{Observables}

\subsubsection{The CMB Lensing Convergence Power Spectrum \label{sec:power_spectrum}}
    At the level of Gaussian fluctuations, the information contained in a map of $\kappa$ can be summarised in terms of its angular power spectrum $C_{\ell}^{\kappa\kappa}$.  In the Limber and Born approximations, it is related to the three-dimensional matter power spectrum by
    \begin{equation}\label{eq:cl_kappa_theory}
    C_{\ell}^{\kappa\kappa} \simeq \frac{9 H_0^4 \Omega_\mathrm{m}^2}{4 c^4} \int  \mathrm{d}r \, P_\delta \left(k=\dfrac{\ell+1/2}{r}, z(r)\right) \frac{\left(r_s-r\right)^2}{r_s^2 a^2}
    \end{equation}
    and can be straightforwardly computed using \texttt{CAMB}.

\begin{figure}[htbp]
            \centering
            \includegraphics[width=0.9\textwidth]{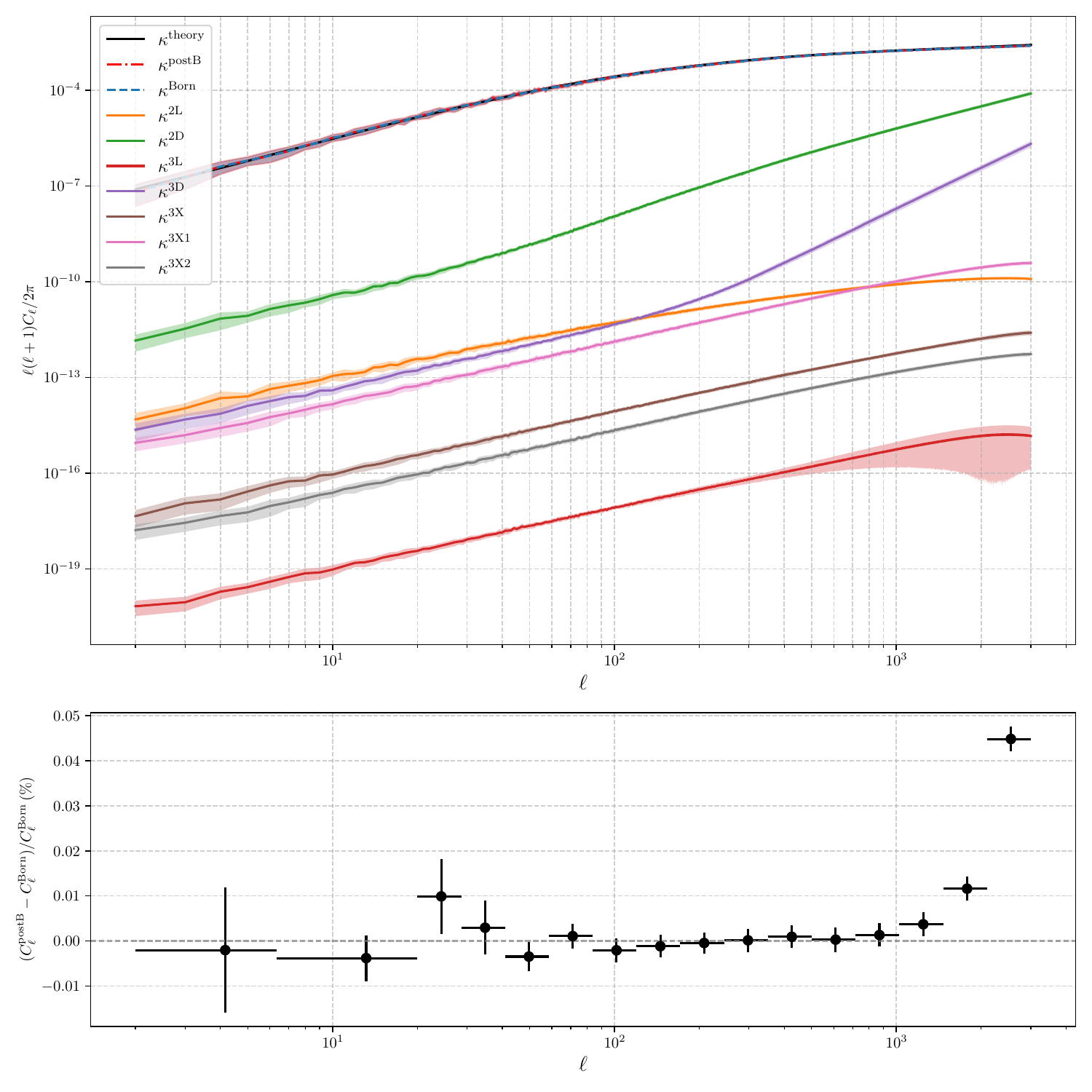}
            \caption{Angular power spectra of the individual Born and post-Born contributions to the lensing convergence, averaged over 30 realisations of our simulated maps (with the error bands denoting the sample standard deviation).
            \label{fig:kappa_angular_power_spectrum}}
        \end{figure}

   
    We show the contributions to the convergence power spectrum from the individual correction terms as well as the total post-Born corrected spectrum in Figure~\ref{fig:kappa_angular_power_spectrum}.  Consistent with the simulation results of Ref.~\cite{Calabrese:2014gla} and the theoretical predictions of Ref.~\cite{Pratten:2016dsm}, we find that the corrections start becoming relevant at multipoles $\ell \gtrsim 1000$.
    
    Among the post-Born terms, the dominant contributions to the power spectrum come from the second-order terms ($2L, 2D$) and two third-order terms ($3D, 3X1$) – omitting the third-order terms ($3L$,$3X$,$3X2$) does not have an appreciable impact on maps or power spectra.  A similar trend is observed in terms of non-Gaussianity, as we will see in Section~\ref{sec:NG}. Therefore, we shall only consider the contributions from the terms ($\kappa^\mathrm{B}, \kappa^{2L}, \kappa^{2D}, \kappa^{3D}, \kappa^{3X1}$) in the following.

\subsubsection{Non-Gaussian Observables}\label{sec:NG}
    We quantify the non-Gaussianity of the lensing convergence map in terms of the dimensionless skewness and kurtosis parameters: 
    \begin{equation}\label{eq:S_para}
	\begin{aligned}
    	&S_0=\frac{\left\langle \kappa^{3}\right\rangle_{\mathrm{c}}}{\sigma_{0}^{3}}, \quad S_1=\frac{\left\langle \kappa|\mathbf{\nabla} \kappa|^{2}\right\rangle_{\mathrm{c}}}{\sigma_{0} \sigma_{1}^{2}} ,\\
    	&S_2=-\frac{\left\langle|\nabla \kappa|^{2} \Delta \kappa\right\rangle_{\mathrm{c}}\sigma_{0}}{\sigma_{1}^{4}},
        \end{aligned}
    \end{equation}
    \begin{equation}\label{eq:K_para}
        \begin{aligned}
    	&K_0=\frac{\left\langle \kappa^{4}\right\rangle_{\mathrm{c}}}{\sigma_{0}^{4}}, \quad K_1=\frac{\left\langle \kappa^{2}|\nabla \kappa|^{2}\right\rangle_{\mathrm{c}}}{\sigma_{0}^2 \sigma_{1}^2}, \\
    	&K_2=-\frac{\left\langle \kappa|\nabla \kappa|^{2} \Delta \kappa\right\rangle_{\mathrm{c}}}{\sigma_{1}^4}, \quad K_3=\frac{\left\langle|\nabla \kappa|^{4}\right\rangle_{\mathrm{c}}}{\sigma_{1}^{4}},
	\end{aligned}
	\end{equation}
    where $\sigma_0 \equiv \left\langle \kappa^2\right\rangle^{1/2}$ and $\sigma_1 = \left\langle (\nabla \kappa)^2\right\rangle^{1/2}$ are the standard deviations of $\kappa$ and of the absolute value of the gradient of $\kappa$.  The $S$- and $K$-parameters can be regarded as summary statistics of the three- and four-point correlations (or the bi- and tri-spectrum) of $\kappa$, respectively.  Unlike the bi- and tri-spectrum, they can easily be numerically evaluated, and can also be used to express non-Gaussian corrections to Minkowski functionals~\cite{Matsubara:2010te}.

    In terms of the non-Gaussianity gain from post-Born corrections, we find a similar contribution pattern as in the power spectrum. As shown in Table~\ref{table:NG-postB}, including the four post-Born correction terms ($B, 2L, 2D, 3D, 3X1$) yields results that are essentially indistinguishable from those obtained when including all second- and third-order post-Born correction terms.

    \begin{longtable}{c|c|c|c|c|c}
        \hline
        \hline
        & \textbf{B} & \textbf{all terms} & \textbf{2 terms} & \textbf{3 terms} & \textbf{4 terms} \\
        \hline
        $\sigma_0\times10^{-2}$ & 6.754 $\pm$ 0.045 & 6.754 $\pm$ 0.045 & 6.779 $\pm$ 0.046 & 6.754 $\pm$ 0.045 & 6.754 $\pm$ 0.045 \\ \hline
        $\sigma_1$ & 97.53 $\pm$ 0.96 & 97.54 $\pm$ 0.96 & 98.37 $\pm$ 0.99 & 97.55 $\pm$ 0.96 & 97.54 $\pm$ 0.96 \\ \hline
        $S_0\times10^{-2}$ & 14.76 $\pm$ 6.57 & 14.95 $\pm$ 6.57 & 14.97 $\pm$ 6.57 & 14.96 $\pm$ 6.57 & 14.95 $\pm$ 6.57 \\ \hline
        $S_1\times10^{-2}$ & 8.753 $\pm$ 3.898 & 8.843 $\pm$ 3.896 & 8.813 $\pm$ 3.881 & 8.845 $\pm$ 3.896 & 8.844 $\pm$ 3.896 \\ \hline
        $S_2\times10^{-2}$ & 4.213 $\pm$ 1.874 & 4.316 $\pm$ 1.873 & 4.274 $\pm$ 1.856 & 4.316 $\pm$ 1.873 & 4.316 $\pm$ 1.873 \\ \hline
        $K_0\times10^{-2}$ & 11.24 $\pm$ 5.01 & 11.43 $\pm$ 5.09 & 11.47 $\pm$ 5.10 & 11.44 $\pm$ 5.10 & 11.43 $\pm$ 5.10 \\ \hline
        $K_1\times10^{-2}$ & 5.028 $\pm$ 2.247 & 5.096 $\pm$ 2.277 & 5.093$\pm$ 2.267 & 5.098 $\pm$ 2.277 & 5.097 $\pm$ 2.277 \\ \hline
        $K_2\times10^{-2}$ & 4.962 $\pm$ 2.219 & 5.020 $\pm$ 2.244 & 5.012 $\pm$ 2.225 & 5.021 $\pm$ 2.244 & 5.020 $\pm$ 2.244 \\ \hline
        $K_3\times10^{-2}$ & 6.431 $\pm$ 2.874 & 6.478 $\pm$ 2.894 & 6.518 $\pm$ 2.874 & 6.480 $\pm$ 2.894 & 6.478 $\pm$ 2.893 \\ 
        \hline
        \hline
        \caption{Gaussian and non-Gaussian parameters for different combinations of Born approximation and post-Born correction terms. The column marked ``all terms'' refers to the combination of all correction terms: B+2L+2D+3D+3X+3L+3X1+3X2; ``2 terms'' refers to B+2L+2D; ``3 terms'' refers to B+2L+2D+3D; and ``4 terms'' refers to B+2L+2D+3D+3X1. The data presented are the mean and standard deviation values obtained from 30 realisations.}\label{table:NG-postB}
    \end{longtable}
    
    The $S$- and $K$-parameters' main downside is that individually, they are insensitive to any scale-dependence of the non-Gaussianity, and in combination only very weakly.  One way to make them more sensitive to scale-dependent non-Gaussianity is by decomposing the convergence map with scale-dependent filters, e.g., spherical needlets~\cite{Narcowich2006LocalizedTF,Marinucci:2007aj} and evaluating the non-Gaussian parameters on the decomposed maps~\cite{Planck:2018lbu, Hamann:2023tdu} – an approach that we will follow here as well.  
    
    We filter convergence maps in terms of spherical harmonic expansion coefficients $a_{\ell m}$, by convolving them with a needlet window function $b$ in harmonic space
    \begin{equation} \label{eq:needlet-decomposition}
        \beta_{j}(\hat{\boldsymbol{n}})=\sum_{\ell=B^{j-1}}^{B^{j+1}} b\left(\frac{\ell}{B^{j}}\right) \sum_{m = -\ell}^{\ell} a_{\ell m} Y_{\ell m}(\hat{\boldsymbol{n}}),
    \end{equation}
    where $\beta_{j}(\hat{\boldsymbol{n}})$ denotes the needlet-filtered map, $j \in \mathbb{N}$ is the scale index, and $B$ is the bandwidth parameter of the needlet. The window function $b$ is supported on $\ell \in \left[B^{j-1}, B^{j+1}\right]$ and satisfies $\sum_{j} b^{2}\left(\frac{\ell}{B^{j}}\right)=1$ for all $\ell$.  
    
\begin{figure}[htbp]
        \centering
        \includegraphics[width=0.84\textwidth]{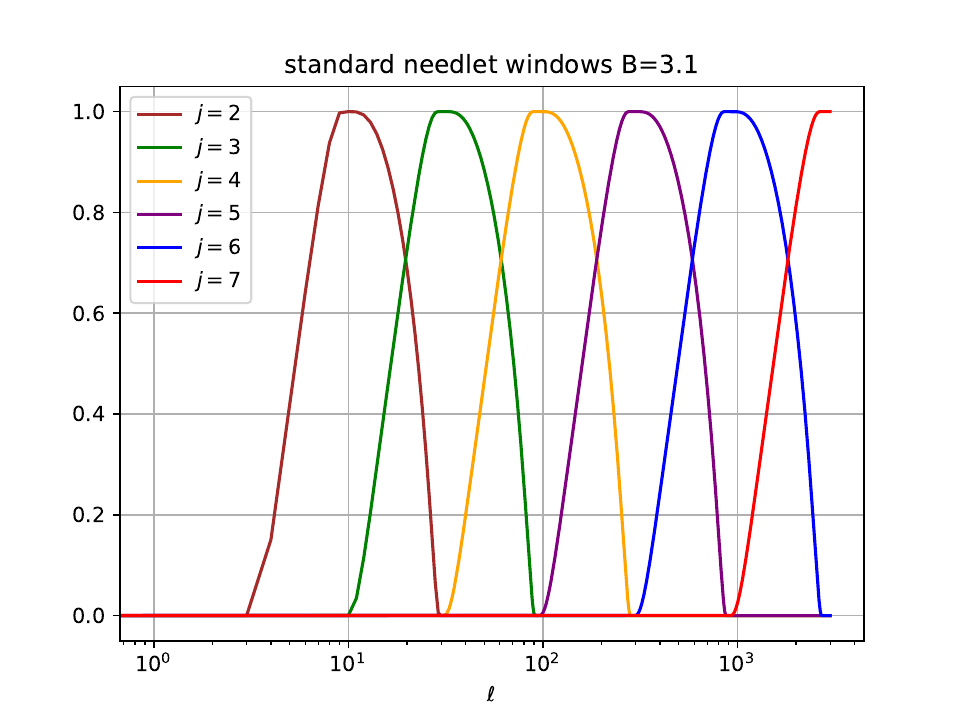}
        \caption{Window functions used for the needlet decomposition, with band width parameter $B = 3.1$ and needlet scale index $j \in \left\{2,3,4,5,6,7 \right\}$. \label{fig:needlet}}
    \end{figure}

    Using the window function construction suggested in Ref.~\cite{Marinucci:2007aj} for a maximum multipole of $\ell = 3000$ and a total of six needlet filters, we choose $B = 3000^{1/7} \simeq 3.1$ and  $j \in \left\{ 2,3,4,5,6,7 \right\}$. This choice ensures that the last filter ($j = 7$) is peaked at $\ell = 3000$, while the first filter ($j = 2$) starts around $\ell = 2$. The corresponding needlet filters are shown in Figure~\ref{fig:needlet}.

    \subsection{Simulation parameter settings}

    \subsubsection{Number of Lensing Shells}\label{sec:Number of Lens Shells}
    Some care must be taken in choosing the number of lenses used in the simulation; if the number is too small, the thin lens approximation becomes a poor one, and the time-dependence of the lens is not properly taken into account either.  Also, given that our simulation pipeline yields uncorrelated lenses, we cannot model correlations of the density field on scales larger than the shell thickness, so very thin shells (i.e., much thinner than the typical scale of radial correlations $\mathcal{O}(100)$~Mpc) should be avoided too.

    \paragraph{Preliminary considerations}  As a preliminary exercise, we investigate the approximate number of lensing shells required within our simulation to obtain a stable result for the Born-approximated $C_{\ell}^{\kappa\kappa}$.  
    For this particular test, we do not use the default setup of lensing shells described in Section~\ref{sec:shell_arrangement} above, but instead for simplicity slice the volume bounded by the surface of last scattering into $N_\mathrm{shell}$ shells of equal comoving distances and generate all density contrast maps as realisations of Gaussian random fields.
    For a range of values of $N_\mathrm{shell}$, we simulate 20 independent realisations of CMB convergence maps each.  
    
\begin{figure}[htbp]
        \centering
        \includegraphics[width=0.8\textwidth]{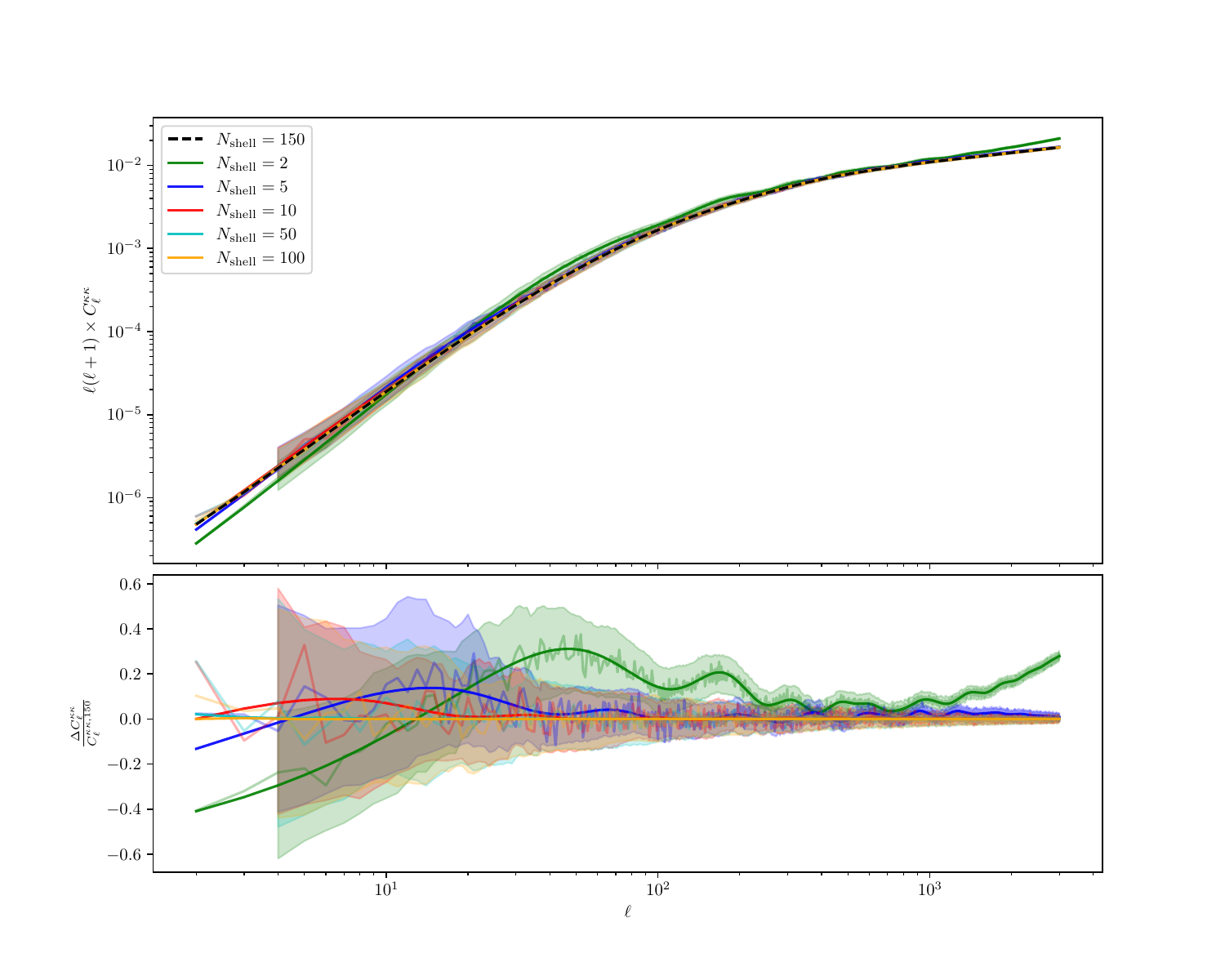}
        \caption{\textit{Upper panel:} Angular power spectrum of the CMB lensing convergence in the Born approximation, averaged over 20 simulated full-sky maps for different numbers $N_\mathrm{shell}$ of lensing shells of equal thickness.  \textit{Lower panel:} relative difference with respect to $N_\mathrm{shell} = 150$.  The simulation means are plotted in semi-transparent thin lines, and thick lines correspond to theoretical expectation values, calculated from Equation~\ref{eq:kappa_numerical}.  Error bands denote the standard deviation of the 20 simulations and have been convolved with top-hat window function of width $\Delta \ell = 5$ for $2 < \ell \leq 30$ and $\Delta \ell = 15$ for $\ell > 30$. 
        \label{fig:val_shell_num}}
    \end{figure}

    The averaged angular power spectra and corresponding standard deviations are plotted in Figure~\ref{fig:val_shell_num} and compared to the $N_\mathrm{shell} = 150$ (i.e., $\Delta r \sim 62~\mathrm{Mpc}/h$) case.  Even with a relatively modest number of lenses $N_\mathrm{shell} \sim 10$, the thin lens approximation yields a spectrum that lies within 10\% of the $N_\mathrm{shell} = 150$ result at large scales $\ell \lesssim 10$ and better than that at smaller scales.  For $N_\mathrm{shell} \gtrsim 50$, the power spectra agree to better than 1\% at all scales. Figure~\ref{fig:val_shell_num} also demonstrates an excellent agreement between the power spectra of the ray-traced simulation maps and the corresponding theoretical expectation values obtained from Equation~\ref{eq:kappa_numerical}.

    \paragraph{A more detailed analysis}
    While it is reassuring that the power spectrum is not strongly dependent on the number of lensing shells, this does not imply that the same must be true for the simulations' non-Gaussian properties.  We therefore revisit the question of number of shells/shell thickness from the perspective of requiring numerical stability for the non-Gaussian parameters of our simulated convergence maps.

\begin{figure}[htbp]
        \centering
        \includegraphics[width=0.8\textwidth]{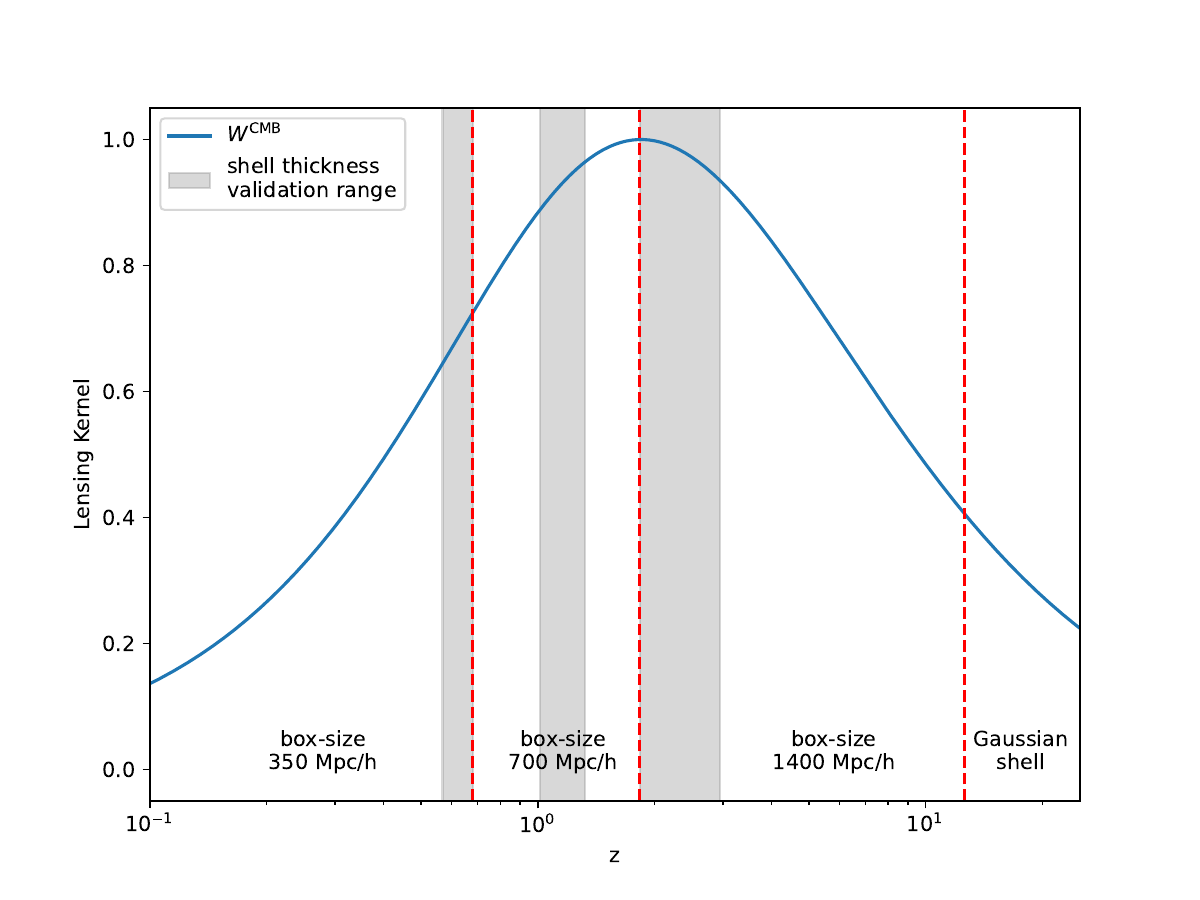}
        \caption{Redshift ranges used for the analysis of shell thickness (gray bands) superimposed on the CMB lensing kernel, $W^{\rm{CMB}}$, with amplitude normalised to one (solid blue line). Vertical dashed red lines indicate the transition redshifts where we change $N$-body box size. \label{fig:shell_thickness_valid_range}}
    \end{figure}

    As the non-Gaussianity of the convergence originates in part from the non-Gaussianity of the lenses, this analysis requires us to use $N$-body generated lensing shells.  However, due to the substantial storage requirements of testing different configurations (namely the need to output and store snapshots at all redshifts $z(r_j)$), we focus on varying the shell thickness in three limited validation ranges – one for each box size – chosen such that they cover redshifts where the lensing kernel is significant (see Figure~\ref{fig:shell_thickness_valid_range}).\footnote{The comoving distances of the validation ranges for each box size are $[1349, 1699] \, \rm{Mpc}/h$, $[2199, 2899] \, \rm{Mpc}/h$ and $[3398, 4798] \, \rm{Mpc}/h]$ for box sizes $350 \, \rm{Mpc}/h$, $700 \, \rm{Mpc}/h$ and $1400 \, \rm{Mpc}/h$, respectively.}  These ranges are then each subdivided into shells of equal thickness.

\begin{figure}[htbp]
        \centering
        \includegraphics[width=1\textwidth]{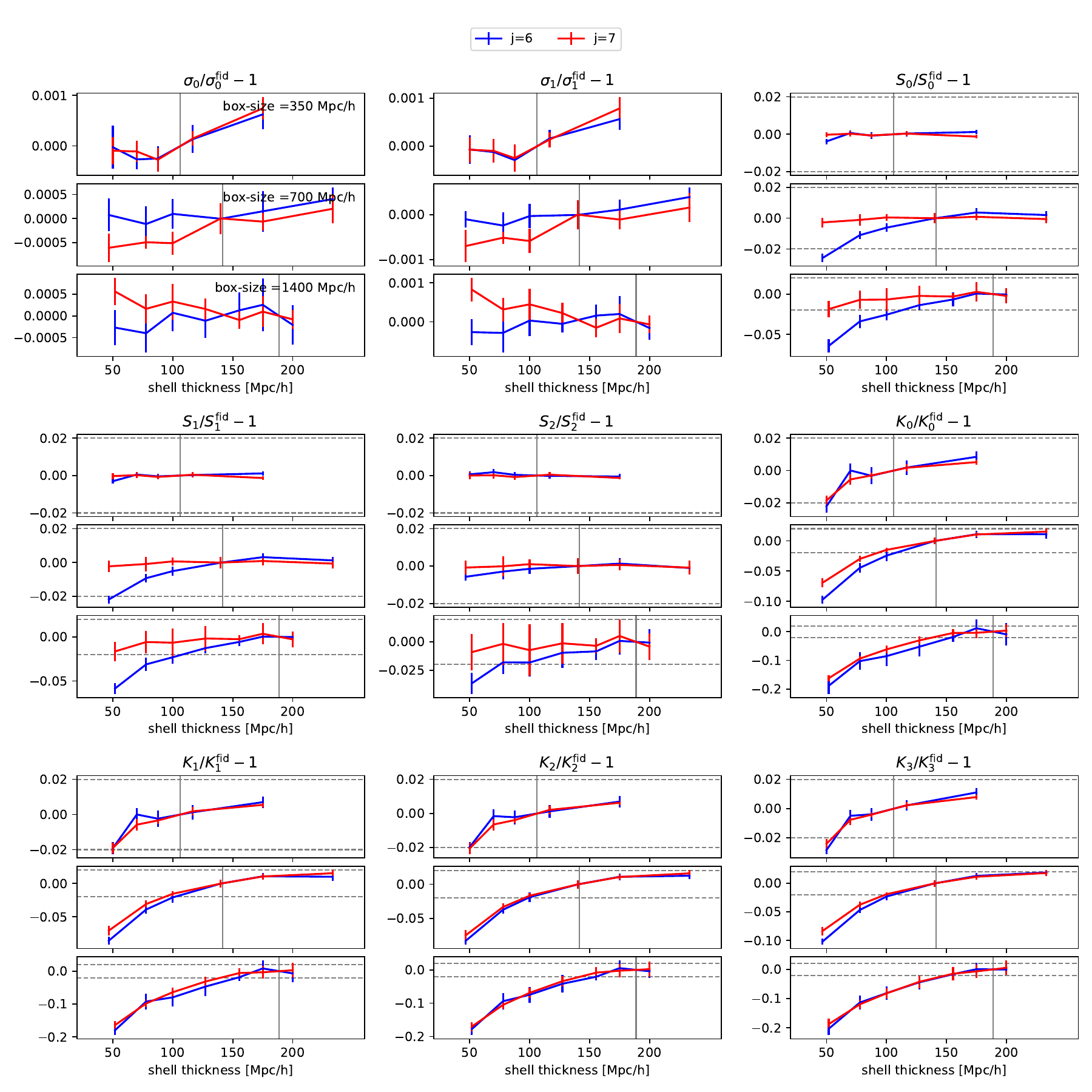}
        \caption{Relative change of $\sigma_0$, $\sigma_1$ and the skewness and kurtosis parameters of needlet-filtered simulated CMB convergence maps as a function of the shell thickness in the validation ranges of the three box sizes.  Error bars indicate the sample standard deviation over 10 realisations.  Blue and red lines correspond to results with needlet parameters $j=6$ and $j= 7$, respectively.  The vertical grey lines denote the shell thickness we adopt for our default configuration.  Horizontal dashed lines mark a variation of $2\%$ around the results of our default settings. \label{fig:shell_thickness_valid}}
    \end{figure}

    Varying the thickness parameters one at a time, we then apply needlet filters to the resulting convergence maps and compute their $\sigma_0$, $\sigma_1$ and the $S$- and $K$-parameters for ten realisations.  The results for needlets with parameters $j = 6$ and $j = 7$ (i.e., on small scales that are less affected by sample variance and where the non-Gaussianities are most significant) are shown in Figure~\ref{fig:shell_thickness_valid}. 

    As expected from the preliminary analysis, the parameters $\sigma_0$ and $\sigma_1$ (which are related to the power spectrum) are largely insensitive to variations in shell thickness between $50 \to 200$~Mpc/$h$, changing only at the per mille level.
    For the smallest box size, there is only a weak dependence of the skewness and kurtosis parameters on shell thickness, and the results converge at the sub-percent level for $\Delta r \gtrsim 100$~Mpc/$h$. The two larger box sizes are somewhat more strongly dependent on the shell thickness and thinner shells $\Delta r \gtrsim 100$~Mpc/$h$ lead to less non-Gaussianity in the convergence map, particularly so for the kurtosis parameters which can be suppressed by up to 20\% (see also \cite{Matilla:2019fxd}).  This may potentially be due to the loss of correlation in the radial direction.  Nonetheless, the gradient levels off at higher $\Delta r$ and there are signs of stabilisation around $\Delta r \sim 150$~Mpc/h – we therefore choose the shell thickness for our default configuration near this region. We estimate that this effect will lead to our predictions for the skewness and kurtosis parameters being subject to a systematic uncertainty of up to 10\% in total, if we take into account potential further contributions from outside our validation ranges.

   \subsubsection{Lens resolution}
\begin{figure}[htbp]
        \centering
        \includegraphics[width=0.84\textwidth]{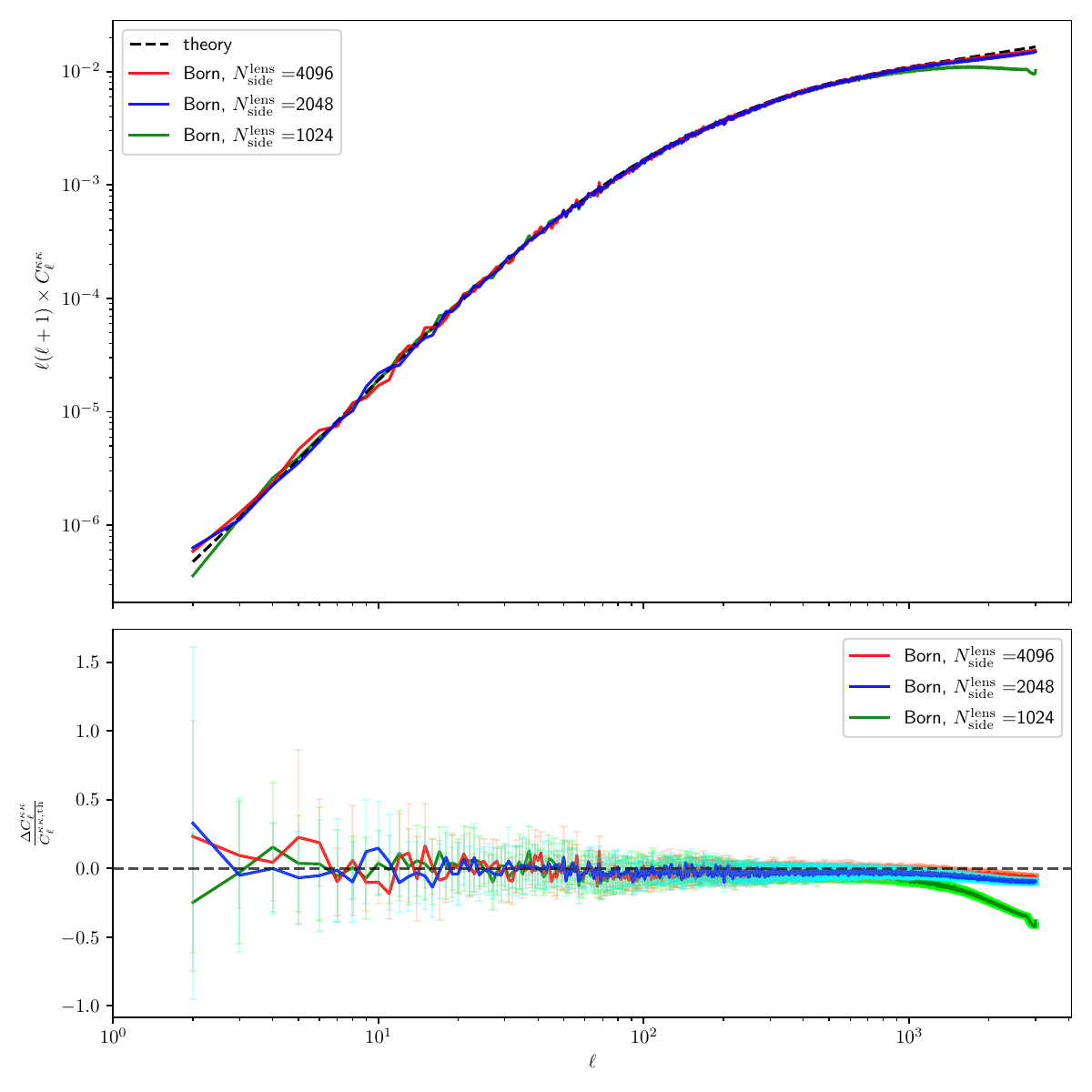}
            \caption{\textit{Upper panel:} Angular power spectrum of the CMB lensing convergence in the Born approximation, averaged over 20 simulated full-sky maps at \texttt{HEALPix} resolutions $N^\mathrm{lens}_{\text{side}}=1024$, $N^\mathrm{lens}_{\text{side}}=2048$ and $N^\mathrm{lens}_{\text{side}}=4096$.  \textit{Lower panel:} relative difference with respect to the \texttt{CAMB} result with 1-$\sigma$ error bars included. \label{fig:convergence_fid_Nside_compare}}
    \end{figure}

   Next we investigate how the \texttt{HEALPix} resolution of the lenses affects the results of our simulations.  Keeping all other parameters fixed to their default values (Table~\ref{tab:shell_settings}), we consider the options $N^\mathrm{lens}_{\text{side}}=1024$, $N^\mathrm{lens}_{\text{side}}=2048$ and $N^\mathrm{lens}_{\text{side}}=4096$.  In Figure~\ref{fig:convergence_fid_Nside_compare}, we show the Born-approximated angular power spectrum of the CMB convergence, averaged over 20 realisations, for these three cases.  Evidently, a resolution of $N^\mathrm{lens}_{\text{side}}=1024$ is not quite sufficient to resolve the power spectrum accurately at $\ell \sim 1000$ and higher, showing a distinct lack of power at small scales.  The spectra for $N^\mathrm{lens}_{\text{side}}=2048$ and $N^\mathrm{lens}_{\text{side}}=4096$ are within a percent of each other all the way up to $\ell = 3000$, suggesting that $N^\mathrm{lens}_{\text{side}}=2048$ may be sufficient here.  
   
   Note that up to $\ell \sim 1000$, there is also very good agreement between simulated $N^\mathrm{lens}_{\text{side}}=2048$ and $N^\mathrm{lens}_{\text{side}}=4096$ spectra and the \texttt{CAMB} prediction, but beyond that the simulation results start to deviate mildly from the \texttt{CAMB} one, reaching about $5\%$ discrepancy at $\ell \sim 2000$ – an effect that was also observed in Ref.~\cite{Takahashi:2017hjr}.  Keeping in mind that we use $N$-body simulations to determine the non-linear growth of structures whereas the \texttt{CAMB} results are derived uses the \texttt{Halofit} model though, this difference is still within the claimed accuracy of \texttt{Halofit}~\cite{Takahashi:2012em}.

    \subsubsection{Redshift of transition from Gaussian to $N$-body shells}
\begin{figure}[htbp]
        \centering
        \begin{minipage}{\textwidth}
        \centering
        \includegraphics[width=0.8\textwidth]{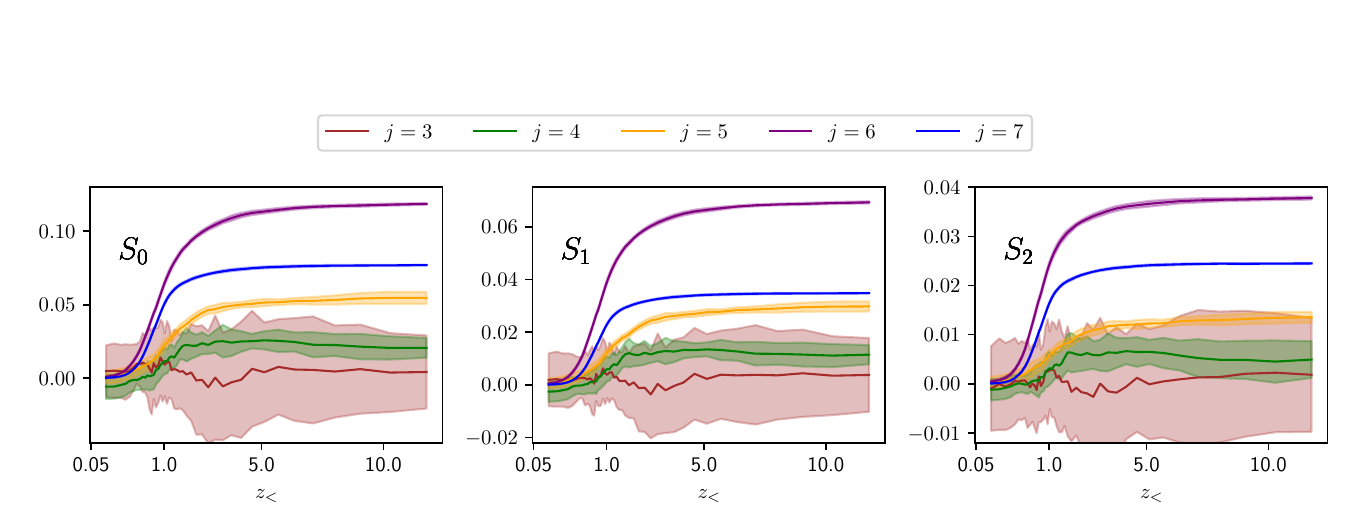}
        \end{minipage}
        \begin{minipage}{\textwidth}
        \centering
        \includegraphics[width=1.05\textwidth]{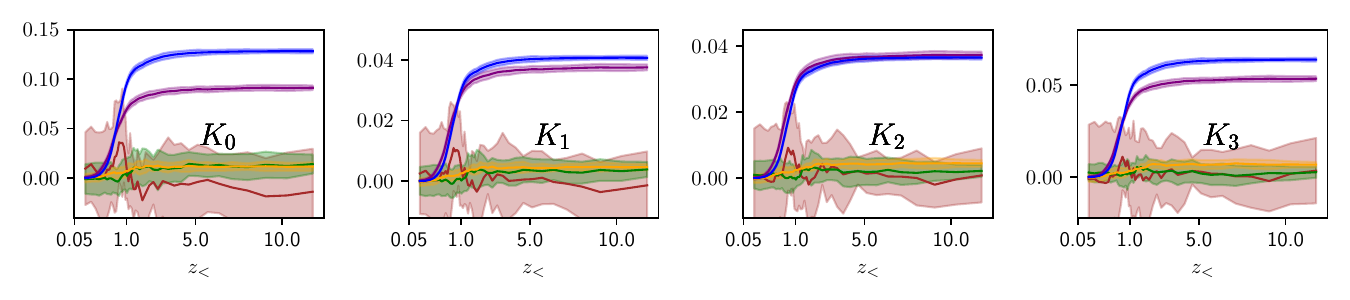}
        \end{minipage}
        \caption{Skewness and kurtosis parameters of needlet-filtered CMB lensing convergence maps as a function of the redshift $z_<$ where the transition from $N$-body shells to Gaussian shells occurs.  Dashed vertical lines indicate where the $N$-body box size changes and error bands represent the sample standard deviation over 10 random realisations of the maps. Note that the scale of the horizontal axis switches from logarithmic to linear at $z_< = 1$. \label{fig:lensing_evolution_valid}}
    \end{figure}

    It is common knowledge that at sufficiently early times, the matter field is well-approximated by a Gaussian random field.  What does ``sufficiently early" mean in our context though?  At what redshift can we safely start using cheap Gaussian realisations instead of expensive $N$-body simulation results?  Starting from our default configuration, we address this question by successively replacing all shells at redshifts $z > z_{<}$ with Gaussian realisations and inspecting the dependence of the non-Gaussian observables on the transition redshift $z_{<}$.  We show the results of this analysis in Figure~\ref{fig:lensing_evolution_valid}. 

    On the largest scales (needlet indices $j < 4$), the sample variance is so large that one cannot expect to extract much useful information from the skewness and kurtosis parameters of a single CMB sky at all.

    For $j \geq 4$, i.e., on intermediate to smaller scales, our simulations clearly predict non-zero values of the non-Gaussianity parameters and the skewness and kurtosis parameters exhibit reasonable numerical convergence around $z_{<} \sim 10$.  The gradient of the curves is a measure of the contribution to the convergence map's non-Gaussianity due to the non-Gaussianity of the lenses.  For the skewness parameters, this contribution is predominantly generated at redshifts $z \sim 2$, roughly coinciding with the peak of the lensing kernel, whereas their contribution to the kurtosis parameters is predominantly generated at lower redshifts $z \lesssim 1$, when the amplitude of density fluctuations is larger and evidently compensates for the suppression by the lensing kernel.  

    Note that the there is a residual non-Gaussianity, even as $z_< \to 0$, i.e., if all lenses are Gaussian.  This originates from the inherent non-linearity of General Relativity and in our case specifically is due to the post-Born corrections.\footnote{Using only the Born approximation and entirely Gaussian lenses yields convergence maps that are consistent with zero skewness and kurtosis.}  While the post-Born corrections make up only an $\mathcal{O}(10\%)$ proportion of the convergence map's skewness, its kurtosis is in fact dominated by them.

\section{Cosmological parameters from CMB lensing non-Gaussianities}\label{sec:constrain}

Let us now turn to the question of how much useful cosmological information is contained in the non-Gaussianity of the CMB convergence.  As a basic estimate, we will consider here the following scenario:
\begin{itemize}
\item{A cosmological parameter space spanned by the cold dark matter density $\Omega_\mathrm{c} h^2$ and the amplitude of the power spectrum of primordial curvature perturbations, $\vec{\Theta} = ( \Omega_\mathrm{c} h^2, A_\mathrm{s})$.
\item{Observables given by $\vec{\mathfrak{O}} = ({\mathfrak{O}}^{(j)}) = ((\sigma_0^{(j)}, \sigma_1^{(j)}, S_0^{(j)}, S_1^{(j)}, S_2^{(j)}, K_0^{(j)}, K_1^{(j)}, K_2^{(j)}, K_3^{(j)}))$}, where the superscript $(j)$ denotes that the corresponding quantities are evaluated on a needlet-filtered map with scale parameter $j$.  The parameters $\sigma_0$ and $\sigma_1$ are representative of the Gaussian information in the convergence map (and in our simple example with only a two-dimensional space of cosmological parameters will have similar constraining power to the full power spectrum), and the skewness and kurtosis parameters represent the non-Gaussian information.}
\item{We do not take into account any reconstruction noise of convergence map and instead use a multipole cutoff, considering two scenarios: a CMB-S4-like experiment, where the convergence map will be signal-dominated up to around $\ell_\mathrm{max} = 1000$~\cite{CMB-S4:2016ple} (this range is covered by $j \in \{2, 3, 4, 5, 6\}$), as well as a more idealistic case with $\ell_\mathrm{max} = 3000$, which requires $j \in \{2, 3, 4, 5, 6, 7\}$ for full coverage.}
\end{itemize}
We use a Fisher matrix analysis~\cite{Fisher:1935:CDJ} to evaluate the constraining power, with the Fisher information matrix given by
\begin{equation}\label{eq:Fisher_matrix_G}
    \mathfrak{F}_{ij} = \frac{\partial \mathfrak{O}_k}{\partial \Theta_i} \frac{\partial \mathfrak{O}_l}{\partial \Theta_j} \Sigma^{-1}_{kl}.
\end{equation}

The covariance matrix $\Sigma$ is constructed from 200 random realisations of our default configuration. Note that the na{\"i}ve estimate
\begin{equation}
    \hat{\Sigma}_{kl} = \left\langle \left(\mathfrak{O}_{k}^{\text{sim}}-\left\langle \mathfrak{O}_{k}^{\text{sim}}\right\rangle\right)\left(\mathfrak{O}_{l}^{\text{sim}}-\left\langle \mathfrak{O}_{l}^{\text{sim}}\right\rangle\right) \right\rangle 
\end{equation}
underestimates the covariance, but as long as $N_\mathrm{sim} > p+2$, where $p$ is the rank of $\Sigma$, the bias can be corrected~\cite{Hartlap:2006kj} via  
    \begin{equation}
    	\Sigma^{-1} = \frac{N_\mathrm{sim}-p-2}{N_\mathrm{sim}-2} \, \hat{\Sigma}^{-1}.
    \end{equation}
    
In order to compute the derivatives $\frac{\partial \mathfrak{O}}{\partial \Theta}$ we generate three additional sets of $N$-body simulations 
for each of four points in parameter space, $\vec{\Theta} = (0.11, 2.119 \times 10^{-9})$, $(0.13, 2.119 \times 10^{-9})$, $(0.1203, 1.7 \times 10^{-9})$ and $(0.1203, 2.54 \times 10^{-9})$, and then average over the results of ten lensing simulations.  We ensure that we use identical random seeds for all stochastic processes ($N$-body initial conditions, selection of $N$-body simulation for each shell, random shell rotations) to suppress the effect of sample variance.

The validity of the Fisher matrix approximation rests on the assumption that the posterior is approximately Gaussian. We confirmed this by verifying that (i) the skewness and kurtosis parameters are roughly linear functions of the cosmological parameters over the range considered, and (ii) that in our simulations, their histograms are close to Gaussian. 

The overall constraining power can be quantified in terms of a figure of merit (FoM)~\cite{Albrecht:2006um} 

\begin{equation}
    \rm{FoM} = \sqrt{\rm{det}(\mathfrak{F})},
\end{equation}
which is inversely proportional to the area of the respective 1-$\sigma$ constraint ellipse for the cosmological parameters.

\begin{figure}[htbp]
    \centering
    \includegraphics[width=0.8\textwidth]{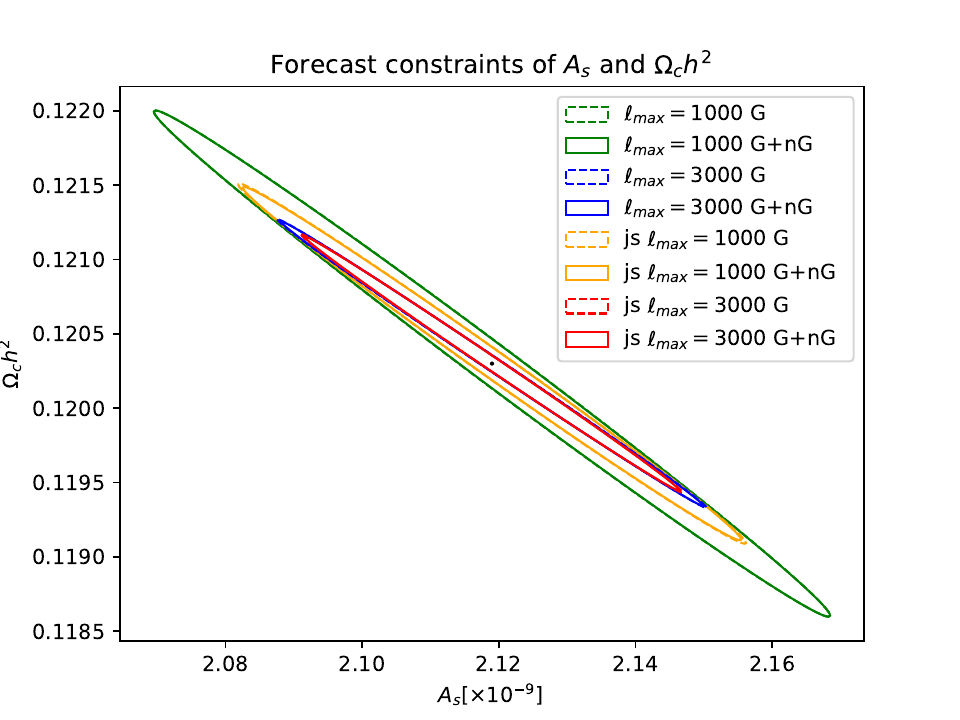}
    \caption{Forecast of 1-$\sigma$ constraints on $\Omega_{\rm{c}} h^2$ and $A_\mathrm{s}$ from analyses with no filtering and joint needlet filtering. Solid and dashed lines correspond to results considering both Gaussian and non-Gaussian parameters and only Gaussian parameters, respectively.\label{fig:constraints}}
\end{figure}

We compare constraints containing purely Gaussian information (“G”), based solely on the parameters $\sigma^{(j)}_0$ and $\sigma^{(j)}_1$, with constraints that combine Gaussian with non-Gaussian information (“G+nG”), which include all nine parameters: $\sigma^{(j)}_0$, $\sigma^{(j)}_1$, as well as the skewness and kurtosis parameters. Our forecasted 1-$\sigma$ constraints are shown in Figure~\ref{fig:constraints} and the corresponding Figures of Merit presented in Table~\ref{tab:FoM}.

\begin{longtable}{c|c|c|c}
    \hline
    \hline
    & \textbf{G}  & \textbf{G+nG} & \textbf{Improvement} 
    \\
    \hline
    no filter ($\ell_{\rm{max}}=1000$)  & 277,785  & 281,398  & 1.3\%  
\\ \hline
    no filter ($\ell_{\rm{max}}=3000$)  & 1,284,656  & 1,331,128  & 3.6\%
\\ \hline
    joint $j$s   & 1,478,465  & 1,543,448  & 4.4\% 
    \\ \hline
    \hline
    \hline
    \caption{Figures of Merit for joint needlet constraints on $\Omega_c h^2$ and $A_\mathrm{s}$. We show the relative improvement in the FoM from adding non-Gaussian information to the Gaussian information.\label{tab:FoM}}
\end{longtable}

The information gain from adding the non-Gaussian parameters to the analysis is fairly modest, increasing the FoM by just over one per cent for $\ell_\mathrm{max} = 1000$, though going to higher resolutions or adding sensitivity to scale-dependent effects via needlet decomposition of the maps can increase this to a few per cent.

\section{Conclusion}\label{sec:conclusion}

Since the first release of \textit{Planck} data~\cite{Planck:2013mth}, measurements of the weak lensing of the CMB have become a powerful probe of cosmology, proving their usefulness not only in combination with other data sets, e.g., helping to break degeneracies in CMB temperature and polarisation data, particularly in extended models, but even when used on their own.  With observations from high-resolution Stage III~\cite{ACT:2023kun,SPT:2023jql,SimonsObservatory:2018koc} and Stage IV~\cite{CMB-S4:2016ple} CMB polarisation experiments, we can expect even more precise insights.  In particular, we may become able to access non-Gaussian information in CMB lensing observables.

However, extracting this information will require similarly precise theoretical predictions – a non-trivial task due to the inherently non-linear nature of gravitational lensing.  In this work, we have developed an approach based on combining ray-tracing methods with $N$-body simulations and linear theory results to address this problem.  Our simulation software, dubbed \texttt{FLAReS} (Full-sky Lensing with Adaptive Resolution Shells), is made freely available for download from  \href{https://github.com/Kang-Yuqi/FLAReS}{GitHub}.\footnote{\href{https://github.com/Kang-Yuqi/FLAReS}{\texttt{https://github.com/Kang-Yuqi/FLAReS}}}  It is relatively inexpensive computationally and applicable to gravitational lensing of the CMB as well as of galaxies.

Using \texttt{FLAReS}, we conducted a suite of CMB lensing simulations to predict the skewness and kurtosis properties of lensing convergence maps, and performed a forecast in order to determine to what extent the non-Gaussian information can help improve constraints on cosmological parameters.  

We found that including non-Gaussian information can improve sensitivity in the $(\Omega_\mathrm{c} h^2, A_\mathrm{s})$-plane by $\sim 4\%$, compared to using only the Gaussian information for a measurement of the convergence map that is signal-dominated up to $\ell_\mathrm{max} = 3000$. The improvement appears rather modest, but two points should be kept in mind when interpreting this number.  Firstly, our approach does not capture radial correlations on scales larger than the shell thickness – which would lead to additional contributions to the non-Gaussianity via lens-lens coupling, and thus potentially enhance the constraining power of the non-Gaussian observables.  Secondly, in larger parameter spaces (e.g., extensions of the base $\Lambda$CDM model), Gaussian observables may be subject to more serious parameter degeneracies that non-Gaussian information may be able to break.  Further exploration of these items will be left to future work.

\acknowledgments
Some of the results in this paper have been derived using the \texttt{healpy} and \texttt{CAMB} packages. This research includes computations using the computational cluster \textit{Katana} supported by Research Technology Services at UNSW Sydney~\cite{Katana}. We thank Antony Lewis for a helpful comment on an earlier version of this paper.

\bibliographystyle{apsrev}
\bibliography{full_sky.bib}

\end{document}